\documentclass[a4paper]{amsart}
\RequirePackage{amsmath}
\RequirePackage{bm}
\RequirePackage{amssymb}
\RequirePackage{upref}
\RequirePackage{amsthm}
\RequirePackage{enumerate}
\RequirePackage{pb-diagram}
\RequirePackage{amsfonts}
\RequirePackage[mathscr]{eucal}
\RequirePackage{verbatim}
\RequirePackage{xr}
\RequirePackage{graphicx}
\usepackage{calc}
\usepackage{xspace}
\RequirePackage{color}
\RequirePackage{ifthen}
\RequirePackage{fancybox}
\RequirePackage{mathrsfs}

\newcommand{\cf}{cf.\@\xspace}
\newcommand{\resp}{resp.\@\xspace}

\newcommand{\al}{\alpha}
\newcommand{\bet}{\beta}
\newcommand{\ga}{\gamma}
\newcommand{\de}{\delta }

\newcommand{\f}{\varphi}
\newcommand{\h}{\eta}

\newcommand{\ka}{\kappa}
\newcommand{\lam}{\lambda}

\newcommand{\om}{\omega}

\newcommand{\s}{\sigma}

\newcommand{\D}{\varDelta}
\newcommand{\F}{\varPhi}
\newcommand{\Lam}{\varLambda}

\newcommand{\so}{{\mc S_0}}
\newcommand{\rhom}{\rho_{\tup{dm}}}
\newcommand{\rhoe}{\rho_{\tup{de}}}

\newcommand{\const}{\tup{const}}

\newcommand{\msp[1]}[1]{\mspace{#1mu}}

\newcommand{\R}[1][n+1]{{\protect\mathbb R}^{#1}}

\newcommand{\Hh}[1][n+1]{{\protect\mathbb H}^{#1}}

\newcommand{\Cc}{{\protect\mathbb C}}

\newcommand{\N}{{\protect\mathbb N}}

\newcommand{\eR}{\stackrel{\lower1ex \hbox{\rule{6.5pt}{0.5pt}}}{\msp[3]\R[]}}
\newcommand{\eN}{\stackrel{\lower1ex \hbox{\rule{6.5pt}{0.5pt}}}{\msp[1]\N}}
\newcommand{\eO}{\stackrel{\lower1ex \hbox{\rule{6pt}{0.5pt}}}{\msc O}}

\DeclareMathOperator{\tr}{tr}

\newcommand\ra{\rightarrow}

\newcommand\pde[2]{\frac {\partial#1}{\partial#2}}
\newcommand{\un}{\infty}
\newcommand{\A}{\forall}

\newcommand{\uu}{\cup}

\newcommand{\uuu}{\bigcup}

\newcommand{\uud}{ \stackrel{\lower 1ex \hbox {.}}{\uu}}
\newcommand{\uuud}[1]{ \stackrel{\lower 1ex \hbox {.}}{\uuu_{#1}}}

\newcommand{\sminus}[1][28]{\raise 0.#1ex\hbox{$\scriptstyle\setminus$}}

\newcommand{\abs}[1]{\lvert#1\rvert}

\newcommand{\spd}[2]{\protect\langle #1,#2\protect\rangle}

\newcommand{\tup}{\textup}% text upright

\newcommand{\mc}{\protect\mathcal}
\newcommand{\msc}{\protect\mathscr}

\providecommand{\bysame}{\makebox[3em]{\hrulefill}\thinspace}

\newcommand{\cq}[1]{\glqq{#1}\grqq\,}

\newcommand{\bt}{\begin{thm}}
\newcommand{\bl}{\begin{lem}}
\newcommand{\bc}{\begin{cor}}
\newcommand{\bd}{\begin{definition}}
\newcommand{\bpp}{\begin{prop}}
\newcommand{\br}{\begin{rem}}
\newcommand{\bn}{\begin{note}}
\newcommand{\be}{\begin{ex}}
\newcommand{\bes}{\begin{exs}}
\newcommand{\bb}{\begin{example}}
\newcommand{\bbs}{\begin{examples}}
\newcommand{\ba}{\begin{axiom}}
\newcommand{\bas}{\begin{assumption}}

\newcommand{\et}{\end{thm}}
\newcommand{\el}{\end{lem}}
\newcommand{\ec}{\end{cor}}
\newcommand{\ed}{\end{definition}}
\newcommand{\epp}{\end{prop}}
\newcommand{\er}{\end{rem}}
\newcommand{\en}{\end{note}}
\newcommand{\ee}{\end{ex}}
\newcommand{\ees}{\end{exs}}
\newcommand{\eb}{\end{example}}
\newcommand{\ebs}{\end{examples}}
\newcommand{\ea}{\end{axiom}}
\newcommand{\eas}{\end{assumption}}

\newcommand{\bp}{\begin{proof}}
\newcommand{\ep}{\end{proof}}
\newcommand{\eps}{\renewcommand{\qed}{}\end{proof}}

\newcommand{\bal}{\begin{align}}

\newcommand{\bi}[1][1.]{\begin{enumerate}[\upshape #1]}
\newcommand{\bia}[1][(1)]{\begin{enumerate}[\upshape #1]}
\newcommand{\bin}[1][1]{\begin{enumerate}[\upshape\bfseries #1]}
\newcommand{\bir}[1][(i)]{\begin{enumerate}[\upshape #1]}
\newcommand{\bic}[1][(i)]{\begin{enumerate}[\upshape\hspace{2\cma}#1]}
\newcommand{\bis}[2][1.]{\begin{enumerate}[\upshape\hspace{#2\parindent}#1]}
\newcommand{\ei}{\end{enumerate}}

\newcommand\ndots{\raise 0.47ex \hbox {,}\hskip0.06em\cdots %
     \raise 0.47ex \hbox {,}\hskip0.06em} 

\newcommand{\q}{\quad}
\newcommand{\qq}{\qquad}

\newcommand\nd{\noindent}

\newskip\Csmallskipamount                                                
\Csmallskipamount=\smallskipamount
\newskip\Cmedskipamount
\Cmedskipamount=\medskipamount
\newskip\Cbigskipamount
\Cbigskipamount=\bigskipamount

\newcommand\cvs{\vspace\Csmallskipamount}   
\newcommand\cvm{\vspace\Cmedskipamount}

\newskip\csa
\csa=\smallskipamount

\newskip\cma
\cma=\medskipamount

\newskip\cba
\cba=\bigskipamount

\newdimen\spt
\newcommand\citem{\cvs\advance\itemno by
1{(\romannumeral\the\itemno})\hskip3pt}
\newcommand{\bitem}{\cvm\nd\advance\itemno by
1{\bf\the\itemno}\hspace{\cma}}

\newcount\itemno
\newcommand{\las}[1]{\label{S:#1}}

\newcommand{\lae}[1]{\label{E:#1}}
\newcommand{\lat}[1]{\label{T:#1}}
\newcommand{\lal}[1]{\label{L:#1}}

\newcommand{\rs}[1]{Section~\ref{S:#1}}

\newcommand{\rl}[1]{Lemma~\ref{L:#1}}

\newcommand{\re}[1]{\eqref{E:#1}}

\newcommand{\fre}[1]{\eqref{E:#1} on page~\tup{\pageref{E:#1}}}

\newcommand{\frs}[1]{Section~\ref{S:#1} on page~\tup{\pageref{S:#1}}}

\newskip\thmskip
\thmskip=\parindent

\newskip\hsk
\newenvironment{hinw}{\labelsep=0pt\begin{list}{}{\labelsep=0pt\itemindent=0pt\labelwidth=0pt\leftmargin=\parindent\rightmargin=0pt\partopsep=\cba}%
\item\it\nopagebreak\nopagebreak}%
{\end{list}}

\newcommand\bh{\begin{hinw}}
\newcommand{\eh}{\end{hinw}}

\newtheoremstyle{normal}% name
  {\cba}%      Space above, empty = `usual value'
  {\cba}%      Space below
  {}% Body font
  {\thmskip}%Indent amount (empty = no indent, \parindent = para indent)
  {\bfseries}% Thm head font
  {.}%        Punctuation after thm head
  {\hsk}%     Space after thm head: " " = normal interword space;
  {}% Thm head spec

\newtheoremstyle{abschnitt}% name
  {\cba}%      Space above, empty = `usual value'
  {\cba}%      Space below
  {}% Body font
  {\thmskip}% Indent amount (empty = no indent, \parindent = para indent)
  {\bfseries}% Thm head font
  {.}%        Punctuation after thm head
  {\hsk}%     Space after thm head: " " = normal interword space;
  {}% Thm head spec

\newtheoremstyle{italic}% name
  {\cba}%      Space above, empty = `usual value'
  {\cba}%      Space below
  {\itshape}% Body font
  {\thmskip}%  Indent amount (empty = no indent, \parindent = para indent)
  {\bfseries}% Thm head font
  {.}%        Punctuation after thm head
  {\hsk}%     Space after thm head: " " = normal interword space;
  {}% Thm head spec

\newtheoremstyle{aufgaben}% name
  {\cba}%      Space above, empty = `usual value'
  {\cba}%      Space below
  {}% Body font
  {}%         Indent amount (empty = no indent, \parindent = para indent)
  {\normalsize\bfseries}% Thm head font
  {.}%        Punctuation after thm head
  {\hsk}%     Space after thm head: " " = normal interword space;
  {}% Thm head spec

\newtheoremstyle{break}% name
  {\cba}%      Space above, empty = `usual value'
  {\cba}%      Space below
  {\itshape}% Body font
  {}%         Indent amount (empty = no indent, \parindent = para indent)
  {\bfseries}% Thm head font
  {.}%        Punctuation after thm head
  {\newline}% Space after thm head: \newline = linebreak
  {}%         Thm head spec

\theoremstyle{italic}
\newtheorem{thm}[subsection]{Theorem}
\newtheorem{lem}[subsection]{Lemma}
\newtheorem{prop}[subsection]{Proposition}
\newtheorem{cor}[subsection]{Corollary}

\theoremstyle{normal}
\newtheorem{rem}[subsection]{Remark}
\newtheorem{definition}[subsection]{Definition}
\newtheorem{example}[subsection]{Example}
\newtheorem{examples}[subsection]{Examples}
\newtheorem{ex}[subsection]{Exercise}
\newtheorem{note}[subsection]{}
\newtheorem{axiom}[subsection]{Axiom}
\newtheorem{assumption}[subsection]{Assumption}

\theoremstyle{aufgaben}
\newtheorem{exs}[subsection]{Exercises}

\numberwithin{equation}{section}
\numberwithin{figure}{section}

\newenvironment{textequation}[1][0.8]
{\begin{equation}
\begin{aligned}
\begin{minipage}{#1\linewidth}}
{\end{minipage}
\end{aligned}
\end{equation}
\ignorespacesafterend}

\newcommand{\btext}{\begin{textequation}}
\newcommand{\etext}{\end{textequation}}

\def\hinweis{\@startsection{subsection}{2}%
 \z@{0.7\linespacing\@plus 0.5\linespacing}{0.7\linespacing}%
{\normalfont\itshape\indent}}

\newcounter{hours}\newcounter{minutes}
\newcommand{\printtime}{%
\setcounter{hours}{\time/60}%
\setcounter{minutes}{\time-\value{hours}*60}%
\ifthenelse{\value{minutes}<10}{\thehours :0\theminutes}{\thehours:\theminutes}}

\usepackage[german,english]{babel}
\usepackage{graphicx}
\RequirePackage{amsmath}
\RequirePackage{bm}
\RequirePackage{amssymb}
\RequirePackage{upref}
\RequirePackage{amsthm}
\RequirePackage{enumerate}
\RequirePackage{pb-diagram}
\RequirePackage{amsfonts}
\RequirePackage[mathscr]{eucal}
\RequirePackage{verbatim}
\RequirePackage{xr}
\RequirePackage{graphicx}
\RequirePackage{mathrsfs}
\usepackage{calc}
\usepackage{xspace}
\usepackage{dsfont}

\makeatletter
\RequirePackage{color}
\newcommand{\ann}[1]{\renewcommand{\@makefnmark}{\mbox{$^{\color{red}{\@thefnmark}}$}}%
\footnote {#1}}
\makeatother

\RequirePackage{upref}
\RequirePackage{amsthm}
\RequirePackage{enumerate}%\begin{enumerate}[(i)]
\usepackage[mathscr]{eucal}
\usepackage{xr-hyper}

\usepackage{calc}

\newlength{\oddsidemarginlength}
\newlength{\topmarginlength}

\newcounter{numberoflines}
\newcounter{tempcc}
\oddsidemargin=\oddsidemarginlength
\evensidemargin=\oddsidemargin
\topmargin=\topmarginlength
\usepackage[colorlinks=true,linkcolor=blue,citecolor=blue,urlcolor=blue]{hyperref}  

\begin{document}

\flushbottom

%\larger[1]
%\frontmatter

\title{Applications of canonical quantum gravity to cosmology}

% author one information
\author{Claus Gerhardt}
\address{Ruprecht-Karls-Universit\"at, Institut f\"ur Angewandte Mathematik,
Im Neuenheimer Feld 205, 69120 Heidelberg, Germany}
%\curraddr{}
\email{\href{mailto:gerhardt@math.uni-heidelberg.de}{gerhardt@math.uni-heidelberg.de}}
\urladdr{\href{http://www.math.uni-heidelberg.de/studinfo/gerhardt/}{http://www.math.uni-heidelberg.de/studinfo/gerhardt/}}
%\thanks{This work was supported by the DFG}

% author two information
%\author{}
%\address{}
%\curraddr{}
%\email{}
%\thanks{}
%
%\subjclass[2000]{35J60, 53C21, 53C44, 53C50, 58J05}
%\keywords{Lorentzian manifold, mass, cosmological spacetime, general relativity, inverse mean curvature flow, ARW spacetimes}


\date{\today}
%
% at present the "communicated by" line appears only in ERA and PROC
%\commby{}

%\dedicatory{}

\begin{abstract} 
We apply quantum gravitational results to spatially unbounded Friedmann universes and try to answer some questions related to dark energy, dark matter, inflation and the missing antimatter.
\end{abstract}

\maketitle

\tableofcontents

\setcounter{section}{0}
%\section{Introduction}
\section{Introduction}
The quantization of gravity is one of the most challenging open problems in physics. The Einstein equations are the Euler-Lagrange equations of the Einstein-Hilbert functional and quantization of a Lagrangian theory requires to switch from a Lagrangian view to a Hamiltonian view. In a ground breaking paper, Arnowitt, Deser and Misner \cite{adm:old} expressed the Einstein-Hilbert Lagrangian in a form which allowed to derive a corresponding Hamilton function by applying the Legendre transformation. However, since the Einstein-Hilbert Lagrangian is singular, the Hamiltonian description of gravity is only correct if two additional constraints are satisfied, namely, the Hamilton constraint and the diffeomorphism constraint. Dirac \cite{dirac:lqm} proved how to quantize a constrained Hamiltonian system---at least in principle---and his method has been applied to the Hamiltonian setting of gravity, \cf the paper by DeWitt \cite{dewitt:gravity} and the monographs by Kiefer \cite{kiefer:book} and Thiemann \cite{thiemann:book}.  In the general case, when arbitrary globally hyperbolic spacetime metrics are allowed, the problem turned out to be extremely difficult and solutions could only be found by assuming a high degree of symmetry, \cf, e.g., \cite{cg:qfriedman}.

However, in the papers \cite{cg:qgravity, cg:uqtheory} we developed a model for the quantization of gravity for general hyperbolic spacetimes. In these papers  we eliminated the diffeomorphism constraint by reducing the number of variables and proving that the Euler-Lagrange equations for this special class of metrics were still the full Einstein equations. The Hamiltonian description of the Einstein-Hilbert functional then allowed a canonical quantization. 

We quantized the action by looking at the Wheeler-DeWitt equation in a fiber bundle $E$, where the base space is a Cauchy hypersurface of the spacetime which has been quantized and the elements of the fibers are Riemannian metrics. The fibers of $E$ are equipped with a Lorentzian metric such that they are globally hyperbolic and the transformed Hamiltonian, which is now a hyperbolic operator, $\hat H$, is a normally hyperbolic operator acting only in the fibers.  The Wheeler-DeWitt equation has the form $\hat Hu=0$ with $u\in C^\un(E,\Cc)$ and we defined with the help of the Green's operator a symplectic vector space and a corresponding Weyl system. 

The Wheeler-DeWitt equation seems to be the obvious quantization of the Hamilton condition. However, $\hat H$ acts only in the fibers and not in the base space which is due to the fact that the derivatives are only ordinary covariant derivatives and not functional derivatives, though they are supposed to be functional derivatives, but this property is not really invoked when a functional derivative is applied to $u$, since the result is the same as  applying a partial derivative.

Therefore, we  discarded the Wheeler-DeWitt equation in  the paper \cite{cg:qgravity2b} and also in the monograph \cite{cg:qgravity-book}, and expressed the Hamilton condition differently by looking at the evolution equation of the mean curvature of the foliation hypersurfaces $M(t)$ and implementing the Hamilton condition on the right-hand side of this evolution equation. The left-hand side, a time derivative, we  replaced by the corresponding  Poisson brackets. After canonical quantization the modified evolution equation was transformed to an equation satisfied by operators which acted on functions $u\in C^\un(E,\Cc)$.

Since the Poisson brackets  became a commutator we could now employ the fact that the derivatives are functional derivatives, since we had to differentiate the scalar curvature of a metric when we applied the operator equation to a smooth function and tried to simplify the resulting equation. As a result of the simplification of the commutator action we obtained an elliptic differential operator in the base space, the main part of which was the Laplacian with respect to a fiber element. Here, we considered functions $u$ depending on the variables  $(x,g_{ij})$, where $x$ is a point in the base space $\so$, $x\in\so$, and $g_{ij}$ is an element of the fibers. The fiber metrics have the form
\begin{equation}
g_{ij}=t^\frac4n\s_{ij},
\end{equation}
where $0<t<\un$ is a time-like fiber variable, which is referred to as time, $n\ge 3$, is the dimension of $\so$ and $\s_{ij}$ is a   Riemannian metric, depending only on $x$, subject to the requirement
\begin{equation}\lae{1.2} 
\det\s_{ij}=\det\chi_{ij},
\end{equation}
\cf \cite[equs. (1.4.103) \& (1.4.104), p. 29]{cg:qgravity-book} and also \cite[Remark 1.6.8]{cg:qgravity-book}. The  arbitrary, but fixed,  metric $\chi_{ij}$ in $\so$ had been introduced to transform the densities $\det g_{ij}$ to functions.

On the right-hand side of the evolution equation the interesting term was $H^2$, the square of the mean curvature. It transformed to a second time derivative, the only remaining derivative with respect to a fiber variable, since the differentiations with respect to the other variables canceled each other. The resulting quantized equation is then a wave equation in a globally hyperbolic spacetime 
\begin{equation}
Q=(0,\un)\times \so,
\end{equation}
of the form
\begin{equation}\lae{1.10.2} 
\frac1{32}\frac{n^2}{n-1}\Ddot u
-(n-1) t^{2-\frac4n}\D u-\frac {n}2 t^{2-\frac4n}Ru+nt^2\Lam u=0,
\end{equation}
where $\so$ is a Cauchy hypersurface of the original spacetime and the Laplacian and the scalar curvature $R$ are formed with respect to a metric $\s_{ij}$ satisfying \re{1.2} and $\Lam$ is a cosmological constant. The function $u$ depends on $(x,t,\s_{ij})$.  

Since the  metric $\chi_{ij}$ is also a fiber metric we may choose $\s_{ij}=\chi_{ij}$ and because it is also arbitrary we may set $\chi_{ij}$ to be the original metric of the Cauchy hypersurface $\so$, \cf \cite[Remark 1.6.8 on page 49]{cg:qgravity-book}. The function $u$ then  only depends  on $(t,x)$, $u=u(t,x)$. For a detailed derivation of  equation \re{1.10.2} we refer to \cite[Chapter 1.6]{cg:qgravity-book} or \cite[Section 6]{cg:qgravity2b}.

When $\so$ is a space of constant curvature then the wave equation, considered only for functions $u$ which do not depend on $x$,  is identical to the equation obtained by quantizing the Hamilton constraint in a Friedmann universe without matter but including a cosmological constant, \cf  \cite[Remark 1.6.11 on page 50]{cg:qgravity-book} or \cite[Remark 6.11]{cg:qgravity2b}.

There exist temporal and spatial self-adjoint operators $H_0$ \resp $H_1$ such that the hyperbolic equation is equivalent to  
\begin{equation}
H_0u-H_1u=0,
\end{equation}
where $u=u(t,x)$. The operator $H_0$ is defined by
\begin{equation}
H_0 w=\f_0^{-1}\{-\frac1{32}\frac{n^2}{n-1}\Ddot w-nt^2\Lam w\},
\end{equation}
where $w=w(t)$, $w\in C^\un_c(\R[]_+,\Cc)$, and $\f_0=t^{2-\frac4n}$, while the definition of $H_1$ is given by
\begin{equation}
H_1v=-(n-1)\D v-\frac n2 Rv,
\end{equation}
where $v=v(x)$, $v\in C^\un_c(\so,\Cc)$. More precisely, the operators $H_i$, $i=0,1$, are the corresponding unique self-adjoint extensions of the operators defined above  in the appropriate function spaces.

 Assuming $\Lam<0$ we proved that $H_0$ has a pure point spectrum with positive eigenvalues $\lam_i$, \cf \cite[Chapter 6.2]{cg:qgravity-book}, especially \cite[Theorem 6.2.5 on page 144]{cg:qgravity-book}, while, for $H_1$, it is possible to find  corresponding eigendistributions  for each of the eigenvalues $\lam_i$, if $\so$ is asymptotically Euclidean or if the quantized spacetime is a black hole with a negative cosmological constant, \cf \cite{cg:uf4,cg:qbh,cg:qbh2} or \cite[Chapters 3--5]{cg:qgravity-book}, and also if $\so$ is the hyperbolic space $\so=\Hh[n]$, $n\ge 3$, \cf \frs{5}. 

Let $w_i$, $i\in\N$, be an orthonormal basis for the temporal eigenvalue problems
\begin{equation}
H_0w_i=\lam_iw_i
\end{equation}
and $v_i$ be corresponding smooth eigendistributions for the spatial eigenvalue problems
\begin{equation}
H_1v_i=\lam_iv_i,
\end{equation}
then
\begin{equation}
u_i=w_iv_i
\end{equation}
are special solutions of the wave equation \re{1.10.2}.

The temporal eigenvalues $\lam_i$ all have multiplicity $1$, the spatial eigenvalues are the same eigenvalues, but they may have higher multiplicities. In case of black holes this is caused by very compelling intrinsic mathematical reasons, \cf \cite[Chapter 6.4]{cg:qgravity-book}, but unless there are either convincing intrinsic or extrinsic reasons, like data, we choose the spatial eigenspaces to be one-dimensional, because the spatial eigenvalues belong in general to the continuous spectrum of the spatial Hamiltonian $H_1$. If $\so$ is the Cauchy hypersurface of a Friedmann universe we only considered smooth spherically symmetric spatial eigenfunctions, which also leads to one-dimensional spatial eigenspaces, \cf \cite[Chapter 6.6]{cg:qgravity-book} for the Euclidean case and \frs{5} for the hyperbolic case.

One can then define an abstract Hilbert space $\mc H$ spanned by the $u_i$ and a self-adjoint operator $H$, unitarily equivalent to $H_0$, such that 
\begin{equation}
Hu_i=\lam_i u_i.
\end{equation}
$e^{-\bet H}$ is then of trace class in $\mc H$ for all $\bet>0$ and the canonical extension of $H$ to the corresponding symmetric Fock space $\mc F$, which is still called $H$, shares this property. Hence, we can define the partition function $Z$,
\begin{equation}
Z=\tr e^{-\bet H},
\end{equation}
 the operator density
\begin{equation}
\hat\rho=Z^{-1} e^{-\bet H},\qq\A\, \bet>0,
\end{equation}
the average energy and the von Neumann entropy in $\mc F$. The eigenvectors $u_i$ can also be viewed as to be elements of $\mc F$ and they  are then also eigenvectors of $\hat \rho$. 

In the present paper we want to apply these quantum gravitational results to cosmology by looking at a Friedmann universe 
\begin{equation}
N=I\times \so,
\end{equation}
where $\so$ is a $n$-dimensional  simply connected space of constant curvature $\tilde\ka$,
\begin{equation}
\tilde\ka\in\{0,-1\},
\end{equation}
i.e., $\so$ is either $\R[n]$ or the hyperbolic space $\Hh[n]$, $n\ge 3$. We tried to answer some questions related to dark energy, dark matter, inflation and the missing antimatter. In doing so we shall also show that assuming a negative cosmological constant is not a contradiction to the observational  result of an expanding universe. Usually a positive cosmological constant is supposed to be responsible for the dark energy and dark matter is sometimes explained by  assuming so-called  extended theories of gravity, confer, e.g., the papers \cite{choudhury:dm} and \cite{etg}. In this paper we rely on general relativity combined with some quantum gravitational ingredients.

Let us summarize the main result as a theorem, where $\rhom$ \resp $\rhoe$ refer to the dark matter \resp dark energy densities, which we defined as eigenvalues of the operator density $\hat\rho$ in $\mc F$,  and $\rho_3$ is a conventional density. $Z$ is the partition function, $T>0$ the absolute temperature and $\lam_0>0$ the smallest eigenvalue of the Hamiltonian $H$. 
\bt
Let the cosmological constant $\Lam$,
\begin{equation}
-1<\Lam<0,
\end{equation}
be given and consider the perfect fluid defined by the density
\begin{equation}
\rho=\rhom+\rhoe+\rho_3
\end{equation}
satisfying the assumptions \re{4.23}, \re{4.24}, \re{4.36} and \re{4.37}. Moreover, we suppose that $\bet=T^{-1}$ and the scale factor $a$ are functions depending on $t$.  The initial value problems
\begin{equation}\lae{4.44.1} 
\frac{\Ddot a}{a}=-\frac{\ka^2}{n(n-1)}\{(n-2)\rho+np\}+\frac2{n(n-1)}\Lam
\end{equation} 
and
\begin{equation}\lae{4.45.1}
\dot\bet=-n\frac{\rhom}{\pde{}\bet (\rho_{\tup{dm}}+\rho_{\tup{de}})}a^{-1}\dot a.
\end{equation}
with initial values $(\bet_0,a_0,\dot a_0)$ are then solvable in $I=[t_0,\un)$ provided $\bet_0>0$ is so large that \fre{2.12} as well as
\begin{equation}\lae{4.46.1}
\frac{2\ka^2}{n(n-1)}Z^{-1}\{1-\frac12 (n-2)\al_0e^{-\bet\lam_0}\}+\frac2{n(n-1)}\Lam>0
\end{equation}
are valid at $\bet=\bet_0$ and $a_0>0$ has to be chosen such that after adding
\begin{equation}
-\frac{\ka^2}{n(n-1)}(n(1+\om_3)-2)\ga_3a_0^{-n(1+\om_3)}
\end{equation}
to the left-hand side of  \re{4.46.1} the inequality still remains valid at $\bet=\bet_0$. The initial value $\dot a_0$ is supposed to be positive. The solution $(\bet,a)$ then satisfies
\begin{equation}\lae{4.48.1} 
\dot\bet>0,
\end{equation}
\begin{equation}
\dot a>0,
\end{equation}
\begin{equation}
\Ddot a>0
\end{equation}
and
\begin{equation}\lae{4.51.1}
\frac2{n(n-1)}\ka^2\rho+\frac2{n(n-1)}\Lam-\tilde\ka a^{-2}>0.
\end{equation}
In order that $(\bet,a)$ also satisfies the first Friedmann equation $\dot a_0$ has to be chosen appropriately, namely, such that the first Friedmann equation is valid for $t=t_0$, which is possible, in view of \re{4.51.1}.
\et

\br
Let us also mention that we use (modified) Planck units in this paper, i.e.,
\begin{equation}
c=\ka^2=\hbar=K_B=1,
\end{equation}
where $\ka^2$ is the coupling constant connecting the Einstein tensor with the stress-energy tensor
\begin{equation}
G_{\al\bet}+\Lam \bar g_{\al\bet}=\ka^2 T_{\al\bet}.
\end{equation}

\er

\section{The dark energy density}\las{2}
In \cite[Remark 6.5.5]{cg:qgravity-book} we proposed to use the eigenvalue of the  density operator  $\hat\rho$ with respect to the vacuum vector $\h$, which is $Z^{-1}$, 
\begin{equation}
\hat\rho\h=Z^{-1}\h,
\end{equation}
as the source of dark energy density, and though this eigenvalue is the vacuum, or zero-point, energy and many authors have proposed the vacuum energy to be responsible for the dark energy, these proposals all assumed the cosmological constant to be positive, while we assume $\Lam<0$ because of the spectral resolution of the wave equation, otherwise the temporal Hamiltonian does not have a pure point spectrum. However, if $\Lam<0$ then we have to assure that $Z^{-1}$ dominates $\Lam$ which will only be the case if 
\begin{equation}
T<T_0=T_0(\abs \Lam).
\end{equation}
Note that $Z$ depends on the eigenvalues $\lam_i$ and on 
\begin{equation}
\bet=T^{-1}.
\end{equation}
First, we emphasize that we shall treat 
\begin{equation}\lae{2.4}
\rho_{\tup{de}}=Z^{-1}
\end{equation}
as a constant, i.e., we shall define the perfect fluid stress-energy tensor by
\begin{equation}\lae{2.5}
T_{\al\bet}=-\rho_{\tup{de}}\bar g_{\al\bet}.
\end{equation}

Let $\lam_i>0$, $i\in\N$,  be the eigenvalues of the temporal Hamiltonian $H_0$ for a given $\Lam<0$ and let $\bar\lam_i$ be the eigenvalues for \begin{equation}
\Lam=-1,
\end{equation}
then
\begin{equation}\lae{2.7}
\lam_i=\bar\lam_i \abs\Lam^\frac{n-1}n,
\end{equation}
\cf \cite[Lemma 6.4.9, p. 172]{cg:qgravity-book}, and define the parameter $\tau$ by
\begin{equation}
\tau=\abs\Lam^\frac{n-1}n,
\end{equation}
where we now assume
\begin{equation}
\abs\Lam<1,
\end{equation}
throughout the rest of the paper. We proved in \cite[Theorem 6.5.6, p. 180]{cg:qgravity-book} that
\begin{equation}
\lim_{\tau\ra 0}Z=\un,
\end{equation}
or equivalently, that
\begin{equation}
\lim_{\tau\ra 0}\rho_{\tup{de}}=0.
\end{equation}
However, we shall now derive a more precise estimate of $\rho_{\tup{de}}=Z^{-1}$ involving $\bet$ and $\Lam$.
\bl\lal{2.1}
For any $\Lam$ satisfying $-1<\Lam<0$, there exists exactly one $T_0>0$ such that
\begin{equation}\lae{2.12}
Z^{-1}(\bet)>\abs\Lam\qq\A\, \bet>\bet_0=T_0^{-1},
\end{equation}
where we recall that
\begin{equation}
\bet=T^{-1}.
\end{equation}
\el
\bp
In view of  \re{2.7} we deduce that
\begin{equation}
Z(\bet)\equiv Z(\bet,\lam_i)=\bar Z(\ga,\bar\lam_i)\equiv \bar Z(\ga),
\end{equation}
where
\begin{equation}
\ga=\bet\abs\Lam^\frac{n-1}n.
\end{equation}
From the relations
\begin{equation}\lae{2.16}
0<E=-\pde{\log Z}\bet=\pde{\log Z^{-1}}\bet,
\end{equation}
\cf \cite[equations (6.5.30) and (6.5.32), p. 176]{cg:qgravity-book},
\begin{equation}
\lim_{\bet\ra \un}Z(\bet)=1,
\end{equation}
and
\begin{equation}
\lim_{\bet\ra 0}Z(\bet)=\un,
\end{equation}
\cf \cite[Theorem 6.5.8, p. 181]{cg:qgravity-book}, we then conclude that there exists exactly one $\ga_0$ such that
\begin{equation}
\bar Z^{-1}(\ga_0)=\abs\Lam
\end{equation}
and, furthermore, that
\begin{equation}
\bar Z^{-1}(\ga)>\bar Z^{-1}(\ga_0)\qq\A\, \ga>\ga_0,
\end{equation}
completing the proof of the lemma.
\ep
 Thus, defining the dark energy density by \re{2.4} and \re{2.5}, we immediately deduce:
 \bt\lat{2.2}
 Let $T_0$ be the temperature defined in \rl{2.1} and assume that the temperature $T$ satisfies $T<T_0$, then the dark energy density guarantees that the Friedmann universe with negative cosmological constant $\Lam$,
\begin{equation}
-1<\Lam<0,
\end{equation}
is expanding such that
\begin{equation}
\dot a>0
\end{equation}
as well as
\begin{equation}
\Ddot a>0.
\end{equation}
 \et
\bp
The Friedmann equations for a perfect fluid with energy $\rho$ and pressure $p$ are
\begin{equation}\lae{2.24}
\frac{\dot a^2}{a^2}=\frac 2{n(n-1)}\ka^2\rho+\frac 2{n(n-1)}\Lam-\tilde\ka a^{-2}
\end{equation}
and
\begin{equation}\lae{2.25} 
\frac{\Ddot a}{a}=-\frac{\ka^2}{n(n-1)}\{(n-2)\rho+np\}+\frac2{n(n-1)}\Lam.
\end{equation}
Choosing $\rho=\rho_{\tup{de}}$ we also specified
\begin{equation}
p=-\rho_{\tup{de}}
\end{equation}
yielding
\begin{equation}
\frac{\Ddot a}{a}=\frac{2\ka^2}{n(n-1)}\rho_{\tup{de}}+\frac2{n(n-1)}\Lam.
\end{equation}
Moreover, in our units,
\begin{equation}
\ka^2=1
\end{equation}
and we also only consider space forms satisfying
\begin{equation}
\tilde\ka\le 0,
\end{equation}
hence the theorem is proved in view of \rl{2.1}.
\ep

\section{The inflationary period}
Immediately after the big bang the development of the universe will have to be governed by quantum gravitational forces, i.e., by the eigenfunctions \resp eigendistributions of the corresponding temporal and spatial Hamiltonians, which we have combined to a single Hamiltonian $H$ acting in an abstract separable Hilbert space $\mc H$ spanned by the eigenvectors $u_i$
\begin{equation}
Hu_i=\lam _iu_i,
\end{equation}
where the eigenvalues all have multiplicity $1$,  are ordered
\begin{equation}
0<\lam_0<\lam_1<\cdots
\end{equation}
and converge to infinity
\begin{equation}
\lim_{i\ra\un}\lam_i=\un.
\end{equation}
The dominant energies near the big bang will therefore be the eigenvalues
\begin{equation}
\lam_i=\spd{Hu_i}{u_i}
\end{equation}
for large $i$ and we shall assume, when considering the development of a Friedmann universe, that this development is driven by a perfect fluid
\begin{equation}\lae{3.5}
T_{\al\bet}=-\rho_i\bar g_{\al\bet},
\end{equation}
where
\begin{equation}
\rho_i=\lam_i.
\end{equation}

Looking at the Friedmann equations
\begin{equation}
\frac{\dot a^2}{a^2}=\frac{2\ka^2}{n(n-1)}\rho_i+\frac2{n(n-1)}\Lam-\tilde\ka a^{-2}
\end{equation}
and
\begin{equation}
\frac{\Ddot a}a=\frac{2\ka^2}{n(n-1)}\rho_i+\frac2{n(n-1)}\Lam
\end{equation}
we conclude that the universe is expanding rapidly depending on the eigenvalue $\rho_i=\lam_i$. The corresponding eigenvector, or particle, $u_i$ will decay after some time and produce lower order eigenvectors or maybe particles that can be looked at as matter or radiation satisfying the corresponding equations of state.

After some time the inflationary period will have ended and only the stable ground state $u_0$,
\begin{equation}
Hu_0=\lam_0 u_0,
\end{equation}
together with conventional matter and radiation will be responsible for the further development of the Friedmann universe.  

The eigenvalue $\lam_0$ is of the order $\abs\Lam^\frac {n-1}n$ in view of \fre{2.7}, hence it will dominate $\Lam$ for small values of $\abs\Lam$. 

\section{The dark matter}
Let $\hat\rho$ be the density operator acting in the Fock space $\mc F$,
\begin{equation}
\hat\rho=Z^{-1}e^{-\bet H},
\end{equation}
where we use the same symbol $H$ to denote the self-adjoint operator $H$ in the separable Hilbert space $\mc H$ as well its canonical extension to the corresponding symmetric Fock space $\mc F_+(\mc H)\equiv\mc F$. In \rs{2} we defined the dark energy density $\rho_{\tup{de}}$ by
\begin{equation}
\rho_{\tup{de}}=\spd{\hat\rho\h}\h=Z^{-1}
\end{equation}
and we propose to define the dark matter density by
\begin{equation}
\rho_{\tup{dm}}=\al_0\spd{\hat\rho u_0}{u_0}=\al_0e^{-\bet\lam_0}Z^{-1},
\end{equation}
where $u_0$ is a unit eigenvector of $H$ satisfying
\begin{equation}
H u_0=\lam_0u_0
\end{equation}
and
\begin{equation}\lae{4.5} 
\al_0>1
\end{equation}
an otherwise arbitrary constant. Its presence should guarantee that there exists $\bet_0>0$ such that
\begin{equation}\lae{4.6}
\pde{}\bet (\rho_{\tup{dm}}+\rho_{\tup{de}})<0\qq\A\, \bet\ge \bet_0,
\end{equation}
as we shall now prove:
\bl\lal{4.1}
Let $\al_0$ satisfy \re{4.5} and $\Lam$
\begin{equation}\lae{4.7}
-1<\Lam\le\Lam_0<0,
\end{equation}
then there exists $\bet_0=\bet_0(\al_0,\abs{\Lam_0})$ such that the inequality \re{4.6} is valid.
\el
\bp
In view \fre{2.16} we have 
\begin{equation}\lae{4.8}
\pde{}\bet (\rho_{\tup{dm}}+\rho_{\tup{de}})=-\al_0\lam_0e^{-\bet\lam_0}Z^{-1}+\al_0e^{-\bet\lam_0}Z^{-1}E+Z^{-1}E,
\end{equation}
where 
\begin{equation}
E=\sum_{i=0}^\un\frac{\lam_i}{e^{\bet\lam_i}-1}=\frac {\lam_0}{e^{\bet\lam_0}-1}+\sum_{i=1}^\un\frac{\lam_i}{e^{\bet\lam_i}-1},
\end{equation}
\cf \cite[equ. (6.5.32), p. 176]{cg:qgravity-book} or simply differentiate. Hence, we obtain
\begin{equation}
\begin{aligned}
E e^{\bet\lam_0}&=\frac {\lam_0e^{\bet\lam_0}}{e^{\bet\lam_0}-1}+\sum_{i=1}^\un\frac{\lam_i}{e^{\bet(\lam_i-\lam_0)}-e^{-\bet\lam_0}}\\
&\le \frac {\lam_0e^{\bet\lam_0}}{e^{\bet\lam_0}-1}+\sum_{i=1}^\un\frac{\lam_i}{e^{\bet(\lam_i-\lam_0)}-1}
\end{aligned}
\end{equation}
and we conclude
\begin{equation}\lae{4.11}
\lim_{\bet\ra\un}Ee^{\bet\lam_0}=\lam_0,
\end{equation}
since
\begin{equation}\lae{4.12}
\begin{aligned}
\sum_{i=1}^\un\frac{\lam_i}{e^{\bet(\lam_i-\lam_0)}-1}&=\sum_{i=1}^\un\frac{\lam_i-\lam_0}{e^{\bet(\lam_i-\lam_0)}-1}+\sum_{i=1}^\un\frac{\lam_0}{e^{\bet(\lam_i-\lam_0)}-1}\\
&\le \sum_{i=1}^\un\frac{\mu_i}{e^{\bet\mu_i}-1}+\lam_0(\lam_1-\lam_0)^{-1}\sum_{i=1}^\un\frac{\mu_i}{e^{\bet\mu_i}-1},
\end{aligned}
\end{equation}
where $\mu_i$ is defined by
\begin{equation}
\mu_i=\lam_i-\lam_0\ge \lam_1-\lam_0>0\qq\A\, i\ge 1.
\end{equation}
Thus the right-hand side of \re{4.12} is estimated from above by
\begin{equation}
(1+\lam_0(\lam_1-\lam_0)^{-1})E(\bet,\mu_i)
\end{equation}
and 
\begin{equation}\lae{4.15}
\lim_{\bet\ra\un}E(\bet,\mu_i)=0,
\end{equation}
\cf\cite[equ. (6.5.71), p. 181]{cg:qgravity-book}. Furthermore, we know
\begin{equation}
\lam_0=\bar\lam_0 \abs\Lam^\frac{n-1}n,
\end{equation}
\cf \re{2.7}. Combining these estimates we conclude that there exists
\begin{equation}
\bet_0=\bet_0(\al_0,\abs{\Lam_0})
\end{equation}
such that
\begin{equation}\lae{4.18}
\pde{}\bet (\rho_{\tup{dm}}+\rho_{\tup{de}})\le -\frac{\al_0-1}2\lam_0e^{-\bet\lam_0}Z^{-1}\qq\A\,\bet\ge \bet_0.
\end{equation}
The limits in \re{4.11} and \re{4.15} are also uniform in $\abs\Lam$ because of  \re{4.7}.
\ep

Dark matter is supposed to be dust, i.e., its pressure vanishes, and hence, $\rhom$ cannot be constant which is tantamount to
\begin{equation}
\bet\not\equiv \const,
\end{equation}
since we assume that $\Lam$ is constant. Thus, $\rhoe$ is also not constant, though we still assume that its stress-energy tensor is defined by
\begin{equation}
T_{\al\bet}=-\rhoe\bar g_{\al\bet}.
\end{equation}
Therefore, we can only establish the continuity equation for
\begin{equation}
\rhom+\rhoe
\end{equation}
and not for each density separately. Let a dot or a prime indicate differentiation with respect  to time $t$, then the continuity equation has the form
\begin{equation}\lae{4.22}
(\rhom+\rhoe)'=-n\rhom a^{-1} \dot a,
\end{equation}
because
\begin{equation}\lae{4.23}
p_{\tup {dm}}=0
\end{equation}
and
\begin{equation}\lae{4.24}
p_{\tup{de}}=-\rhoe.
\end{equation}
The left-hand side of \re{4.22} is equal to
\begin{equation}
\pde{}\bet (\rho_{\tup{dm}}+\rho_{\tup{de}})\dot \bet
\end{equation}
and we see that the continuity equation can only be satisfied if
\begin{equation}\lae{4.26}
\dot\bet=-n\frac{\rhom}{\pde{}\bet (\rho_{\tup{dm}}+\rho_{\tup{de}})}a^{-1}\dot a.
\end{equation}
From \rl{4.1} we immediately derive
\bl\lal{4.2}
Let  the assumptions of \rl{4.1} be satisfied and suppose that $\dot a>0$, then, for any solution $\bet=\bet(t)$ of \re{4.26} in the interval
\begin{equation}
I=[t_0,b),\qq t_0<b\le\un,
\end{equation}
with initial value
\begin{equation}
\bet(t_0)\ge\bet_0
\end{equation}
the inequality
\begin{equation}\lae{4.29} 
\dot\bet>0
\end{equation}
is valid and hence
\begin{equation}
\bet(t)\ge\bet_0\qq\A\,t\in I.
\end{equation}
Furthermore, $\dot\bet$ can be expressed in the form
\begin{equation}\lae{4.31}
\dot\bet=n\de(\al_0-1)^{-1}\al_0 a^{-1}\dot a,
\end{equation}
where $\de=\de(t,\bet_0)$ satisfies
\begin{equation}\lae{4.32}
1\le\de\le2
\end{equation}
and
\begin{equation}\lae{4.33} 
\lim_{\bet_0\ra\un}\de=1,
\end{equation}
i.e.,
\begin{equation}\lae{4.34} 
\bet(t)-\bet(t_0)\approx n\de\al_0(\al_0-1)^{-1}(\log a(t)-\log a(t_0)).
\end{equation}
\el
\bp
\cq{\re{4.29}}\q Follows from \re{4.6} and \re{4.26}.

\cq{\re{4.31}}\q To prove the claim we combine \re{4.8}, \re{4.26} and \re{4.11}.

\cq{\re{4.32}} and \cq{\re{4.33}}\q Same argument as before.

\cq{\re{4.34}}\q Obvious in view of \re{4.31} and \re{4.33}.
\ep

Now, we are prepared to solve the Friedmann equations \re{2.24} and \fre{2.25} for 
\begin{equation}
\rho=\rhom+\rhoe+\rho_3,
\end{equation}
where $\rho_3$ is a conventional density satisfying the equation of state
\begin{equation}\lae{4.36}
p_3=\om_3\rho_3
\end{equation}
assuming
\begin{equation}\lae{4.37}
\om_3>-1.
\end{equation}
$\rho_3$ is only added for good measure and we are allowed to assume
\begin{equation}
\rho_3=0,
\end{equation}
since its presence is not essential.

We also emphasize that we have to solve an additional third equation, namely, equation \re{4.26}. We shall solve the Friedmann equations and \re{4.26} in the interval
\begin{equation}
I=[t_0,\un),\qq t_0>0,
\end{equation}
for the unknown functions $(a,\bet)$ with prescribed positive initial values $(a_0,\dot a_0, \bet_0)$. $\bet_0$ can be arbitrary but large enough such that the assumptions in \rl{4.1} and \rl{4.2} are satisfied. If $\rho_3$ vanishes then $a_0>0$ can be arbitrary, otherwise it has to be large enough. The last initial value $\dot a_0>0$ cannot be arbitrary, instead it has to be chosen such that the first Friedmann equation is initially valid  at $t=t_0$.

If these assumptions are satisfied then we shall solve the equations \fre{2.25} and \re{4.26}. The first Friedmann equation will then be valid automatically. For simplicity we shall only consider the case
\begin{equation}
\rho_3>0
\end{equation}
to avoid case distinctions. Then we deduce, from the continuity equation, 
\begin{equation}
\rho_3=\ga_3a^{-n(1+\om_3)},
\end{equation}
where $\ga_3>0$ is a given constant.

Let us now prove:
\bt
Let the cosmological constant $\Lam$,
\begin{equation}
-1<\Lam<0,
\end{equation}
be given and consider the perfect fluid defined by the density
\begin{equation}
\rho=\rhom+\rhoe+\rho_3
\end{equation}
satisfying the assumptions \re{4.23}, \re{4.24}, \re{4.36} and \re{4.37}. Moreover, we suppose that $\bet=T^{-1}$ and the scale factor $a$ are functions depending on $t$.  The initial value problems
\begin{equation}\lae{4.44} 
\frac{\Ddot a}{a}=-\frac{\ka^2}{n(n-1)}\{(n-2)\rho+np\}+\frac2{n(n-1)}\Lam
\end{equation}
and
\begin{equation}\lae{4.45}
\dot\bet=-n\frac{\rhom}{\pde{}\bet (\rho_{\tup{dm}}+\rho_{\tup{de}})}a^{-1}\dot a.
\end{equation}
with initial values $(\bet_0,a_0,\dot a_0)$ are then solvable in $I=[t_0,\un)$ provided $\bet_0>0$ is so large that \fre{2.12} as well as
\begin{equation}\lae{4.46}
\frac{2\ka^2}{n(n-1)}Z^{-1}\{1-\frac12 (n-2)\al_0e^{-\bet\lam_0}\}+\frac2{n(n-1)}\Lam>0
\end{equation}
are valid at $\bet=\bet_0$ and $a_0>0$ has to be chosen such that after adding
\begin{equation}
-\frac{\ka^2}{n(n-1)}(n(1+\om_3)-2)\ga_3a_0^{-n(1+\om_3)}
\end{equation}
to the left hand side of  \re{4.46} the inequality still remains valid at $\bet=\bet_0$. The initial value $\dot a_0$ is supposed to be positive. The solutions $(\bet,a)$ then satisfy
\begin{equation}\lae{4.48} 
\dot\bet>0,
\end{equation}
\begin{equation}
\dot a>0,
\end{equation}
\begin{equation}
\Ddot a>0
\end{equation}
and
\begin{equation}\lae{4.51}
\frac2{n(n-1)}\ka^2\rho+\frac2{n(n-1)}\Lam-\tilde\ka a^{-2}>0.
\end{equation}
In order that $(\bet,a)$ also satisfy the first Friedmann equation $\dot a_0$ has to be chosen appropriately, namely, such that the first Friedmann equation is valid for $t=t_0$, which is possible, in view of \re{4.51}.
\et
\bp
By introducing a new variable
\begin{equation}
\f=\dot a
\end{equation}
we may consider a flow equation for $(\bet, a,\f)$, where $\dot\f$ replaces $\Ddot a$ and
\begin{equation}
\dot a =\f
\end{equation}
is an additional equation.

Choosing then $\bet_0, a_0$ as above and $\f_0>0$ arbitrary the flow has a solution on an maximal time interval
\begin{equation}
I=[t_0,t_1),\qq t_1>t_0,
\end{equation}
because of \rl{4.1} and \rl{4.2}. It is also obvious that the relations \re{4.48}--\re{4.51} are valid, in view of these lemmata.

Furthermore, if the interval $I$ was bounded, then the flow would have a singularity at $t=t_1$ which is not possible, in view of the relation \re{4.34}, which would imply that $\bet$, $\dot\bet$ as well as $a$ and $\dot a$ would tend to infinity by approaching $t_1$ which, however, contradicts the second Friedmann equation  \re{4.44} from which we then would infer
\begin{equation}
0<\Ddot a\le c a\qq\A\, t\in I
\end{equation}
an apparent contradiction. Hence we deduce
\begin{equation}
I=[t_0,\un).
\end{equation}
It remains to prove that the first Friedmann equation is satisfied if $\dot a_0$ is chosen appropriately, Define
\begin{equation}
\F=\dot a^2-\{\frac 2{n(n-1)}\ka^2\rho+\frac 2{n(n-1)}\Lam\}a^2+\tilde\ka,
\end{equation}
then we obtain
\begin{equation}
\dot\F=0,
\end{equation}
in view of the continuity equations and \re{4.44}, yielding
\begin{equation}
\F(t)=\F(t_0)=0\qq\A\, t\in I.
\end{equation}
\ep

\section{The missing antimatter}
In \cite[Theorem 4.3.1, p. 110]{cg:qgravity-book} we proved that a temporal eigenfunction $w=w(t)$ defined in $\R[]_+$ can be naturally extended past the big bang singularity $\{t=0\}$ by defining
\begin{equation}
w(-t)=-w(t),\qq \A\,t>0.
\end{equation}
The extended function is then of class $C^{2,\al}$,
\begin{equation}
w\in C^{2,\al}(\R[]),
\end{equation}
for some $0<\al<1$ and its restriction to $\{t<0\}$ is also a  solution of the variational eigenvalue problem. Hence we have two quantum spacetimes
\begin{equation}
Q_-=\R[]_-\times \so
\end{equation}
and 
\begin{equation}
Q_+=\R[]_+\times \so
\end{equation}
and a $C^{2,\al}$ transition between them. If we assume that the common time function $t$ is future directed in both quantum spacetimes, then the singularity in $\{t=0\}$ would be a big crunch for $Q_-$ and a big bang for $Q_+$ and similarly for the corresponding Friedmann universes $N_\mp$ governed by the Einstein equations. No further singularities will be present, i.e., the spacetime $N_-$ will have no beginning but will end in in a big crunch and will be recreated with a big bang as the spacetime $N_+$.

This scenario would be acceptable if it would describe a cyclical universe. However, there are no further cycles, there would only be one transition from a big crunch to a big bang. Therefore, the  mathematical alternative, namely, that at the big bang two universes with opposite light cones will be created, is more convincing, especially, if the CPT theorem is taken into account which would require that the matter content in the universe with opposite time direction would be antimatter. This second scenario would explain what happened to the missing antimatter.

\section{Spherically symmetric eigenfunctions in hyperbolic space}\las{5}

The spatial Hamiltonian $H_1$ is a linear elliptic operator
\begin{equation}
H_1v=-(n-1)\D v-\frac n2 Rv,
\end{equation}
where the Laplacian is the Laplacian in $\so$ and $R$ the corresponding scalar curvature. We are then looking for eigenfunctions or, more precisely, eigendistributions $v$,
\begin{equation}
H_1v=\lam v,
\end{equation}
such that, for each temporal eigenfunction $(\lam_i,w_i)$ there exists a matching spatial pair $(\lam_i,v_i)$. The product
\begin{equation}
u_i=w_i v_i
\end{equation}
would then be a solution of the wave equation \fre{1.10.2}.

If $\so$ is the hyperbolic space $\Hh[n]$, $n\ge 3$, we have
\begin{equation}
R=-n(n-1)
\end{equation}
and, given any temporal eigenvalue $\lam_i$, we would have to find functions $v_i$ satisfying
\begin{equation}
-(n-1)\D v_i=(\lam_i-\frac{n^2}2(n-1))v_i.
\end{equation}
We are also looking for spherically symmetric eigenfunctions $v_i$. In hyperbolic space the radial eigenfunctions, known as spherical functions, are well-known: For each $\mu\in\Cc$ there exists exactly one radial eigenfunctions $\f_\mu$ of the Laplacian satisfying
\begin{equation}
-\D\f_\mu=(\mu^2+\rho^2)\f_\mu
\end{equation}
and
\begin{equation}
\f_\mu(0)=1,
\end{equation}
where
\begin{equation}
\rho=\frac{n-1}2,
\end{equation}
see e.g., \cite[Section 2]{anker:wave} and the references therein. Here, we introduced geodesic polar coordinates $(r,\xi)$ in $\Hh[n]$ and the $\f_\mu$ only depend on $r$. The $\f_\mu$ have the integral representation
\begin{equation}
\f_\mu(r)=c_n(\sinh r)^{2-n}\int_{-r}^r(\cosh r-\cosh t)^\frac{n-3}2 e^{-i\mu t} dt,
\end{equation}
\cf \cite[equation (6), p. 4]{anker:hyperbolic}. 

Since the $\f_\mu$ are distributions they are smooth in $\Hh[n]$, \cf \cite[Theorem 3.2, p.~125]{lions:book}. Furthermore, for each $i\in\N$ we can choose $\mu_i\in\Cc$ such that
\begin{equation}
(n-1)(\mu_i^2+\rho^2)=\lam_i-\frac{n^2}2(n-1).
\end{equation}
Obviously, there are two solutions $\mu_i$ and $-\mu_i$, but the corresponding eigenfunctions are identical as can be easily checked.

%\backmatter
%\includepdf[pages=-]{/Users/claus/Documents/Scanned-Documents/}
\bibliographystyle{hamsplain}
%\bibliography{mrabbrev,publications}
\providecommand{\bysame}{\leavevmode\hbox to3em{\hrulefill}\thinspace}
\providecommand{\href}[2]{#2}

%\listoffigures

%\cleardoublepage

%\thispagestyle{empty}
%\closegraphsfile
\end{document}